\documentclass[aps,preprint]{revtex4-1}
\usepackage{graphicx}
\usepackage{amssymb, amsmath, amsthm, amsfonts, bm, mathrsfs, wasysym, txfonts, bbm, enumerate, color}
\usepackage{slashed} 

\begin{document}

\title{{\large Gauged Nambu-Jona-Lasinio model and axionic QCD string}}

\author{Chi Xiong}
\email[]{xiongchi@ntu.edu.sg}
\affiliation{Institute of Advanced Studies, Nanyang Technological University, Singapore 639673 }


\begin{abstract}
We propose an axionic QCD string scenario based on the original flux-tube model by Kogut and Susskind, and then incorporate it  into a gauged Nambu-Jona-Lasinio (NJL) model. Axial anomaly is studied by a new topological coupling from the string side, and by the 't Hooft vortex from the NJL side, respectively.  The nontrivial phase distribution of the quark condensate plays an important role in this scenario. 

Keywords: NJL model, QCD string, flux tube, Strong CP problem.

\end{abstract}

\maketitle              

\section{Introduction}

The Nambu-Jona-Lasinio (NJL) model and the flux-tube model are two phenomenological models of quantum chromodynamics (QCD).  The NJL model can handle very well the symmetry feature of QCD (see e.g. Refs. \cite{Hatsuda, Klevansky} for review) --- there are a lot of symmetries and some of them are broken in different ways. One of important symmetry issues is the breaking of chiral symmetry. The NJL model enjoys phenomenological success for hadrons. As shown in Ref. \cite{Hatsuda}, the NJL model agrees to all the empirical facts within a 10-20 \% level. Nevertheless, the NJL model also has well-known shortcomings. Firstly it is not a renormalizable theory, hence, one has to take care of the cutoff issue; secondly, it cannot address the confinement problem. This is probably more serious than the non-renormalizability since some relevant physics for quarks and hadrons could be missing.  

The QCD flux-tube model or QCD string, on the other hand, provides a phenomenological description for quark confinement \cite{KS74, Casher}. For mesons, the color electric field lines are collimated into a flux tube connecting the quark and anti-quark pair, resulting in a linear potential. In this picture the confinement problem is to explain why these field lines are collimated, which will not be investigated in the present article. (Also the chiral symmetry breaking does not seem to be relevant in the flux-tube picture.) Instead we incorporate the flux-tube model into the NJL theory. Of course this cannot be a simple medley of two phenomenological models. Our approach is based on the Cho-Faddeev-Niemi decomposition of the gauge fields of QCD \cite{Cho, Faddeev, DuanGe, Cho99, Kondo08}, which is quite convenient in selecting an Abelian sector of QCD. 

What do we obtain from combining these two models?  We will show that the QCD string can be made axionic (its meaning will be clarified in the next section), and the 't Hooft vertex, or the 't Hooft determinant term \cite{'tHooft:1976, Schafer:1996} could be obtained from the flux-tube configuration, hence the $U_A (1)$ problem can be studied and solved in the quark sector. If QCD with a $\theta$-term is the fundamental theory of the NJL model, it is hard to imagine that such a small, dimensionless parameter $\theta < 10^{-10}$ could play a role in determining those NJL couplings. 

In this paper we follow the same strategy as in Ref. \cite{XC, XC1}. The $\theta$ parameter is replaced by the phase of the quark condensate $\alpha(x)$, which is topologically nontrivial due to the existence of axionic QCD string. Applying the anomaly-inflow mechanism \cite{Callan-Harvey, XC} to the string configuration, we found that it is possible to obtain the  't Hooft vertex from the NJL side and a topological coupling $\partial_{\mu} \alpha K^{\mu} $, which can be considered as a superfluid current $\partial_{\mu} \alpha$ coupled to a Chern-Simons current $K^{\mu}$, from the QCD-string side. Technically we first modify the original flux-tube model\cite{KS74} so it can describe an axionic QCD string. We then construct a gauged NJL model based on this string model and QCD. A new scheme of solving the strong CP problem is proposed in the last section. 

\section{Gauged NJL and axionic QCD string}

We start with a system consisting of a complex scalar $\phi$, an Abelianized gluon potential $Z_{\mu} $ and quark fields $\psi^{ia}$, where the indices $i$ and $a$ are flavor and color indices, respectively.  We will work within the SU(2) QCD and two flavors ($N_f = 2$) for simplicity. The complex scalar 
\begin{equation}
\phi \equiv \textrm{Re}\phi + i \, \textrm{Im}\phi = |\phi| \,e^{i \alpha} 
\end{equation}
is related to the quark condensate as
\begin{equation}
\left\langle \bar{\psi}^i_R \psi^j_L  \right\rangle = - \phi \, \delta^{ij} =  - |\phi| \,e^{i \alpha} \, \delta^{ij}
\end{equation}  
which would be real if the vacuum has a definite parity. However, we will show the importance of introducing the phase field $\alpha = \alpha (x)$. Note that the quark condensates have different values, e.g. $\left\langle \bar{u} u \right\rangle = - (245 \, \textrm{MeV})^3$, while $\left\langle \bar{s} s \right\rangle \approx 0.78 \left\langle \bar{u} u \right\rangle$ \cite{Hatsuda}. In our case $N_f =2$ we simply use one value; in $N_f > 2$ cases there may be more than one phenomenological scalar field, in which case we may regard the $\phi$ field as a column matrix with different scalar fields as its components.

The symmetry group of the NJL model that we will study is 
${SU}_V(2) \otimes {SU}_A(2) \otimes U_V(1) \otimes U_A(1) $.
Note that under the $U_A(1)$ transformation the quarks transform as 
\begin{equation}
\psi_L \rightarrow e^{ - i \xi /2} \psi, ~~~\psi_R \rightarrow e^{ i \xi /2} \psi_R, 
\end{equation}
hence the scalar fields $\phi$ and $\phi^*$ transform as
\begin{equation}
\phi \rightarrow e^{- i \xi} \phi, ~~~ \phi^* \rightarrow e^{ i \xi} \phi^*
\end{equation}
The abelianized gluon potential $Z_\mu$ comes from the Cho-Faddeev-Niemi decomposition (CFN) \cite{Cho, Faddeev, DuanGe, Cho99, Kondo08}. There could be other abelianizing approaches and we choose the CFN decomposition since it can reveal the Abelian sector of QCD and the associated topological defects in a gauge-independent way. A detailed explanation will be given later in this section.

With the system $(\phi, Z_\mu, \psi)$ we will construct a generalization of the original flux-tube model by Kogut and Susskind \cite{KS74}. As in \cite{KS74} we introduce a dielectric function of the vacuum, $\chi(\phi, \phi^*)$ and an effective potential $V(\phi, \phi^*)$ whose expression will be given later. The exact form of the function $\chi$ is not important, e.g. could be a fourth-order polynomial in $(\phi \phi^*)^{1/2}$ as in Ref. \cite{KS74}. It is required that $\chi \rightarrow 0$ when $\phi \rightarrow 0$ and $\chi \rightarrow 1$ for large $|\phi|$.
The Lagrangian is written as 
\begin{eqnarray} \label{preNJL}
\mathcal{L}_{\textrm{eff}} &=& -\frac{1}{4} \chi(\phi, \phi^*) ~Z_{\mu\nu} Z^{\mu\nu} + \partial_{\mu} \phi \partial^{\mu} \phi^* - V(\phi, \phi^*)  - j_{\mu} J^{\mu}  - j_{\mu} K^{\mu}  \cr
&+& \bar{\psi} \big[ i \gamma^{\mu}(\partial_{\mu}  - i g Z_{\mu}) - g_Y ( \textrm{Re}\phi +  i \gamma^5 \textrm{Im}\phi ) + \cdots \big]\, \psi. 
\end{eqnarray}
The function $\chi(\phi, \phi^*)$ describes the color electric and magnetic polarization properties of the vacuum as a physical medium. The currents 
\begin{equation}
j_{\mu} \equiv  -\frac{i}{2 |\phi|^2}\left( \phi^{\ast}\partial_{\mu}\phi-\phi\partial_{\mu}\phi^{\ast}\right) =  \partial_{\mu} \alpha
\end{equation}
and $J_\mu$ represents some external current. We also include a topological current
\begin{equation}
K^{\mu} = \epsilon^{\mu\nu\rho\tau} Z_{\nu} Z_{\rho\tau}
\end{equation}
which is the Chern-Simons current. Note that if the $\phi$ field is used as the order parameter for describing superfluid, the current $j_{\mu}$ is actually related to the superfluid velocity. One may wonder if the phase of the $\phi$ field can be removed away by some rotation of the quark phases. This cannot be done if the $(\phi, Z_\mu, \psi)$ system has nontrivial topological configurations. The phase of the $\phi$ field could be quite complicated (see the Fig. 3 in Ref. \cite{XC2} for an example of the phase distribution of a vortex lattice) and such a rotation does not exist due to the topological obstruction. Nevertheless if the phase distribution is topologically trivial it is possible for the $\phi$ field to be real.
In that case and if we also set $g_Y =0$, the current $j_\mu$ vanishes and the couplings $ j_{\mu} J^{\mu}$ and  $ j_{\mu} K^{\mu}$ drop out and the Lagrangian (\ref{preNJL}) reduces to 
\begin{equation} \label{KS}
\mathcal{L}_{\textrm{KS}} = -\frac{1}{4} \chi(\phi) ~Z_{\mu\nu} Z^{\mu\nu} + \frac{1}{2} \partial_{\mu} \phi \partial^{\mu} \phi - V(\phi)  + \bar{\psi} \big[ i \gamma^{\mu}(\partial_{\mu}  - i g Z_{\mu})  \big]\, \psi 
\end{equation}
which is the original flux-tube model constructed by Kogut and Susskind in Ref. \cite{KS74}, where they show that dynamical nonlinearities, described by the scalar field $\phi$ and the associated dielectric and potential functions,  can trap electric flux lines into tube-like configurations (see \cite{KS74} for details). We stress on that our flux-tube, or axionic QCD string, is different from \cite{KS74} in the following aspects:
\begin{itemize}
\item[1)] We use a complex field $\phi$ whose magnitude is related to the quark condensate; 
\item[2)] The gradient of the phase of $\phi$ is coupled to a topological current (Chern-Simons current);
\item[3)] The complex field has a Yukawa-type coupling to the quark fields;
\item[4)] The Abelian gauge field comes directly from QCD -- Abelianized gluon potential from the CFN decomposition;
\end{itemize}
plus other insignificant differences (like the $j_\mu J^\mu$ coupling to some external current). We have explained 1) at the beginning of this section; About 3), it makes the string ``axionic" (similar to the Callan-Harvey axion-string \cite{Callan-Harvey}); Now we show how 4) proceeds and then move on to 2) which plays a very important role in our model.    

The CFN decomposition splits the gluon potential $A_\mu$ into two new gauge potentials   
\begin{equation} \label{CFN}
A_{\mu} = \tilde{A}_{\mu} + B_{\mu}
\end{equation}
where the new variables $\tilde{A}_{\mu}$ and $B_{\mu}$ are defined as
\begin{equation}
\tilde{A}_{\mu} = (A_{\mu} \cdot {\bf n}) {\bf n} + i g^{-1} [ {\bf n}, \partial_{\mu} {\bf n} ], ~~~
B_{\mu} = i g^{-1} [ \nabla_{\mu} {\bf n}, {\bf n}]
\end{equation}
with $\nabla_{\mu} {\bf n} = \partial_{\mu} {\bf n} - i \, g [A_{\mu}, {\bf n}]$ and a unit vector ${\bf n}$ in the color space. Under the gauge transformation $ \tilde{A}_{\mu}$ transforms like the original gauge field $A_{\mu}$, while $B_{\mu}$ and ${\bf n}$ transform covariantly in the adjoint representation. 
The Abelian character of the potential $\tilde{A}_{\mu}$ can be shown manifestly by introducing  a basis $({\bf n}^1, {\bf n}^2, {\bf n}^3)$ in the SU(2) color space \cite{Cho99}, plus a chromoelectric potential  
\begin{equation}
C_{\mu} = A_{\mu} \cdot {\bf n}^3
\end{equation}
 and a chromomagnetic potential
\begin{equation}
H^a_{\mu} \equiv - \frac{1}{2g} \epsilon^{abc}  {\bf n}^b \partial_{\mu} {\bf n}^c
\end{equation}
Then the gauge potential $\tilde{A}_{\mu}$ can be rewritten as
\begin{equation} \label{U1}
\tilde{A}_{\mu}  = \Omega^{\textrm{\tiny{vac}}}_{\mu} + Z_{\mu} {\bf n}^3
\end{equation}
where $\Omega^{\textrm{\tiny{vac}}}_{\mu}$ represents a classical QCD vacuum since the corresponding field strength vanishes, i.e. $\Omega_{\mu\nu}^{\textrm{\tiny{vac}}}=0$ and $Z_{\mu} $ is the Abelianized gluon potential that we look for.
\begin{equation}
\Omega^{\textrm{\tiny{vac}}}_{\mu} \equiv - H^a_{\mu} {\bf n}^a, ~~~Z_{\mu} \equiv C_{\mu} + H^3_{\mu}
\end{equation}
Then it is easy to see that the field strength $\tilde{F}_{\mu\nu}$ 
\begin{equation} \label{CHmunu}
\tilde{F}_{\mu\nu} \equiv   \partial_{\mu} \tilde{A}_{\nu} - \partial_{\nu} \tilde{A}_{\mu} - i \,g [\tilde{A}_{\mu}, \tilde{A}_{\nu}]= Z_{\mu\nu} \,{\bf n}^3,~~~~ Z_{\mu\nu}= \partial_\mu Z_\nu - \partial_\nu Z_\mu
\end{equation}
has an Abelian feature as expected \cite{Cho99}. Now the Yang-Mills action becomes
\begin{equation} \label{YM}
S_{\textrm{YM}} = \int d^4 x \big[ - \frac{1}{4} Z_{\mu\nu}^2  - \frac{1}{4} B_{\mu\nu}^2 -\frac{1}{2} B^{\mu} Q_{\mu\nu} B^{\nu} \big]   
\end{equation}
where the kinetic term of the Abelianized gluon potential $Z_\mu$ has been separated from the other part, and $Q_{\mu\nu}$ is an operator with second-order derivatives and mixes the $Z_\mu$ and $B_\mu$.  The quark sector of the QCD action in terms of the new variables becomes 
\begin{equation}
S_{\textrm{quark}} = \int d^4x [ \bar{\psi} ( i \gamma^{\mu}(\partial_{\mu}  - i g \tilde{A}_{\mu})  - \mathcal{M}_Q ) \psi + g  \bar{\psi} \gamma^{\mu} t_a  \psi B_{\mu}^a ].
\end{equation}
Integrating out $B_{\mu}$ yields the {\it gauged} NJL effective action \cite{Kondo08, XC}
\begin{equation} 
S_{\textrm{\tiny{gNJL}}} =  \int d^4x \bigg(\bar{\psi} [ i \gamma^{\mu}(\partial_{\mu}  - i g \tilde{A}_{\mu}) - \mathcal{M}_Q ] \psi +\int d^4 y \,G(y) [ \bar{\psi}(x+y) \Gamma_{A} \psi(x - y)\,\bar{\psi}(x-y) \Gamma_{A} \psi(x + y)] \bigg) 
\end{equation}
where $\Gamma_A$ are matrices in Dirac, color and flavor spaces. The double integrals reflect the non-local feature of the four-fermion couplings. However, here we are only interested in a local NJL model. This can be readily done by taking $G(x) = G \, \delta(x)$ ($G$ is the NJL coupling constant).  After this simplification one obtain  a local, gauged NJL action which can be parametrized as
\begin{equation} \label{eff}
\mathcal{L}_{\textrm{\tiny{eff}}} = \bar{\psi} \big[ i \gamma^{\mu}(\partial_{\mu}  - i g \tilde{A}_{\mu}) - G ( \sigma +  i \gamma^5 \vec{\pi} \cdot \vec{\tau} ) + \cdots \big] \psi  
\end{equation}
where the $\sigma$-terms and $\pi$-terms represent the scalar channel and the pseudoscalar channel respectively as in \cite{Hatsuda}. The ellipsis includes contributions from the vector and pseudovector channels.  
Comparing (\ref{preNJL}) with (\ref{eff}) plus (\ref{YM}), we can find some correspondence except for the topological term $j_{\mu} K^{\mu}$ in (\ref{preNJL}) which will emerge in a more sophisticated way.  
One may obtain meson spectra as collective excitations in the vacuum, and collective excitations are fluctuations of the mean fields. Now the question is, can the mean fields or condensates allow a vortex configuration, as required by an axionic string model? Comparing (\ref{eff}) and (\ref{preNJL}) we have
\begin{equation}
\phi \sim \sigma + i \vec{\pi} \cdot \vec{\tau} = f e^{i \alpha}.
\end{equation}
In order to describe, say a single straight vortex configuration with winding number $m$, one takes the amplitude and the phase of  $\phi$ to be, respectively,
\begin{equation} \label{alpha-theta}
f = f(\rho), ~~~~\alpha = m \theta, ~~ m = \pm 1, \pm 2, \cdots 
\end{equation}
with boundary condition 
\begin{equation} \label{vortexbc}
f (0) = 0, ~  (\infty) = \textrm{constant} .
\end{equation}
Following the standard route of computing the effective potential, one obtains
\begin{eqnarray}
V(\phi, \phi^*)  & \equiv &  \mathcal{V}_{\textrm{eff}} ( \sigma, \pi) \cr  
&=& - \frac{1}{4 G} ( \sigma^2 + \pi^2) - \frac{i }{2} \textrm{Tr} \ln \bigg(  (\partial_{\mu} - i g \tilde{A}_{\mu}) ^2  - \frac{g}{2} \sigma_{\mu\nu} \tilde{F}_{\mu\nu} + \sigma^2 + \pi^2 - i\epsilon \bigg).
\end{eqnarray}
Suganuma and Tatsumi \cite{Suganuma} have studied this type of potential and found a critical strength for the restoration of chiral symmetry
\begin{equation}
E_{\textrm{c}} \approx 4 ~ \textrm{GeV/fm}.
\end{equation}
Ref. \cite{Klevansky} estimated the strength of the color electric field in the flux tube using results from \cite{Casher}
\begin{equation}
E \approx 5.3 ~\textrm{GeV/fm} > E_{\textrm{c}}
\end{equation}
which suggests that chiral symmetry should be restored inside the flux tube or the interior of mesons. Hence the boundary condition (\ref{vortexbc}) can be satisfied and vortex configurations are allowed. 

Now we choose a longitudinal variation of gauge potential with respect to $ \hat{\Omega}^{\textrm{\tiny{vac}}}_{\mu} $ (note that its field strength  $\Omega_{\mu\nu}^{\textrm{\tiny{vac}}}=0$)
\begin{equation} \label{Az}
Z_{\mu} = ( Z_{0}, Z_{1}, 0, 0),
\end{equation}
the Dirac equation becomes \cite{Callan-Harvey}
\begin{eqnarray}  \label{Dirac4D}
i \gamma^i (\partial_i - ig Z_i) \psi_L + i( \gamma^2 \cos \theta  + \gamma^3 \sin \theta ) \partial_{\rho} \psi_L+ f(\rho) e^{- i \theta} \psi_R &=&0, \cr
i \gamma^i (\partial_i - ig Z_i) \psi_R + i( \gamma^2 \cos \theta  + \gamma^3 \sin \theta ) \partial_{\rho} \psi_R + f(\rho) e^{+ i \theta} \psi_L &=& 0, 
\end{eqnarray}
where $i = 0, 1$. The appearance of fields $Z_i$ does not affect the main features of the solution --- it is chiral and has the same exponential profile as in Ref.\cite{Callan-Harvey}
\begin{equation} \label{4Dpsi}
\psi_L = \chi_{L} \, \exp \big[- \int_0^{\rho} f(\rho') d\rho' \big]
\end{equation}
where the two-dimensional spinor $\chi_{L} (x_0, x_1)$ satisfies $i \gamma^i (\partial_i - ig Z_i) \chi_{L} = 0$.
It is interesting to compare this localization of chiral zero modes in the flux-tube with the instanton case. 't Hooft has shown that chiral zero modes are located in the instanton \cite{'tHooft:1976} with a profile $\lambda^{3/2}/[(x-a)^2+\lambda^2 ]^{3/2}$ ($\lambda$ and $a$ are the instanton parameters). Noticing that the instanton configuration leads to an effective fermion interaction vertex, so-called  't Hooft vertex, one may ask whether the flux-tube configuration might lead to a similar effective vertex, due to the localization of chiral zero modes. Let us apply a general method in Ref. \cite{Creutz2007} by including fermionic source terms into the partition functional
\begin{eqnarray}
Z (\eta, \bar{\eta}) &=& \int [\mathcal{D}A]  [\mathcal{D}\bar{\psi}] [\mathcal{D} \psi] e^{ - S_A - (\bar{\psi}, ~(\slashed{D}+m)  \psi) - (\bar{\psi}, ~\eta) - (\bar{\eta},~ \psi)}\cr
 &=& \int [\mathcal{D}A] e^{ - S_A +~ (\bar{\eta}, ~(\slashed{D}+m)^{-1}  \eta)} \prod_i (\lambda_i + m) 
\end{eqnarray}
where $\lambda_i$ are eigenvalues of the operator $\slashed{D}$ in a flux-tube background, and $m$ is a small explicit mass introduced for the infrared issue of massless quarks. Suppose that the quark sources have overlaps with the chiral zero mode, say $(\bar{u} \cdot \psi_{0L}) \neq 0$, then a factor of $m^{-1}$ in the source term cancels a factor of $m$ from the determinant part, so the flux-tube configuration can contribute to the correlation functions. Consequently an effective interaction $\sim (\bar{u} \cdot \psi_{0L}) (  \psi^{\dagger}_{0L} \cdot u) \sim \bar{u} (1 \pm \gamma_5 )u $ will appear (note that the zero modes are chiral). For $N_f \geq 2$ flavors,  each flavor has to contribute for the cancellation of mass factors and because of Fermi statistics, this leads to an effective $2 N_f$ vertex which is in the form of a determinant, i.e. the 't Hooft vertex
\begin{equation}
\mathcal{L}_{\textrm{tHooft}} \sim \textrm{det}[ \bar{\psi_i} (1- \gamma_5) \psi_j] + h.c. 
\end{equation}
which should be added to the NJL effective Lagrangian (\ref{eff}). Then it has no problem in addressing the U(1) problem and produces the correct mass for the $\eta'$ meson \cite{Hatsuda}. 
The 't Hooft vertex was originally derived in the instanton scenario. However Greutz's argument above seems to work as well in the flux-tube case, mainly because localization and chirality of the zero modes in such a configuration. Therefore the flux-tube induced quark interaction adds the 't Hooft vertex term to the NJL action. 

This completes our construction of the NJL effective action based on a QCD string model. Now we need to show the emergence of the topological coupling in the Lagrangian (\ref{preNJL})
\begin{equation} \label{cc}
j_{\mu} K^{\mu} \sim \partial_{\mu} \alpha K^{\mu}.
\end{equation}
 The chiral zero modes are also coupled to the Abelianized gluon potential $Z_{\mu}$, therefore a gauge anomaly appears in the vortex $\mathcal{D}^k J_k = \frac{1}{2 \pi} \epsilon^{ij} \partial_i Z_j, ~(i,j,k = 0,3).$ This can be cancelled by an effective action \cite{Callan-Harvey}
\begin{equation} \label{cseff}
S_{\textrm{\tiny{C-S}}} = -\frac{ g^2 N_f}{16 \pi^2}\int d^4 x \, \partial_{\mu} \theta \, K^{\mu}
\end{equation}
which is exactly the new topological coupling (\ref{cc}) that we proposed ($\alpha =  \theta$ as we are considering a static, straight flux tube). This cancellation happens  because the massive quark modes which live off the vortex mediate an effective interaction between the quark condensate and the gluon field, which induces a vacuum current  \cite{Callan-Harvey}
\begin{equation}
J^{\textrm{\tiny{ind}}}_{\mu} =  \frac{g^2 N_f}{8 \pi^2} \epsilon_{\mu\nu\rho\tau} Z^{\rho\tau} \partial^{\nu} \theta.
\end{equation}
Converting it to an effective action we obtain (\ref{cseff}). Therefore a topological term which reflects the axial anomaly can be derived from the axionic QCD string model, which is not a pure Yang-Mill configuration like the instanton.

\section{Discussions and outlook}

As we have shown two phenomenological models -- the NJL model and the flux tube model, can be combined and studied directly from QCD. We make some assumptions, like assuming that the QCD string is axionic, and some simplifications, like neglecting the non-local feature in deriving the NJL effective Lagrangian. Also we have no idea about how to derive the dielectric function $\chi(\phi)$ from a first principle. These are the limitations of our approach.  What's new is the topological coupling $ j_{\mu} K^{\mu} \sim \partial_{\mu} \alpha K^{\mu} $ and the related physics. Note that this coupling emerges at  the quantum level and is neither gauge-invariant nor CP-invariant. However its gauge-variance is what is exactly needed for cancelling the gauge-anomaly localized in the flux-tube. On the other hand, a constant flux tube configuration seems to be CP-violating, which motivates us to propose that in QCD we should have, schematically,
\begin{eqnarray}
\delta|_{\textrm{\tiny{gauge}}} ( \partial_{\mu}\alpha \, K^{\mu}+ \textrm{flux-tube}) &=& 0, \cr
\delta|_{\textrm{\tiny{CP}}} ( \partial_{\mu}\alpha \, K^{\mu} + \textrm{flux-tube}) &=& 0.
\end{eqnarray}
Moreover, if the quark condensate has a non-trivial phase distribution, it will lead to a lot of interesting phenomenologies which have been observed in condensed matter physics, like those in superfluids, superconductors and Bose-Einstein condensates. In fact superfluid or two-fluid models for quark-gluon plasma have been studied recently in Refs.\cite{Chernodub, Kalaydzhyan}. The Josephson effect in QCD has also been considered long time ago in Ref. \cite{Minkowski} and recently in Ref. \cite{XC1}.

\section*{Acknowledgments}

The author thanks Kerson Huang and Peter Minkowski for valuable discussions. He also thanks the hospitality of the Institute for Theoretical Physics, University of Bern where this work was partially done. The author is supported by the research funds from the Institute of Advanced Studies, Nanyang Technological University, Singapore.

\end{document}